\documentstyle[prl,aps,psfig,multicol]{revtex}
\def\ie{{\it i.e.,}\thinspace}

\def\displayfrac#1#2{\frac{\displaystyle #1}{\displaystyle #2}}
\begin{document}
\title{General radiation states and Bell's inequalities} 
\author{Arvind$^1$\cite{email} and
N. Mukunda$^2$\cite{jakkur}} 
\address{$^1$Department of Physics,
Guru Nanak Dev University, Amritsar 142005, India}
\address{$^2$Center for Theoretical Studies and 
Department of Physics\\
Indian Institute of Science,  Bangalore - 560 012, India}
\maketitle 
\begin{abstract}
The connection between quantum optical nonclassicality and
the violation of Bell's inequalities is explored.  Bell
type inequalities for the electromagnetic field are
formulated for general states(arbitrary number or photons,
pure or mixed) of quantised radiation and their violation
is connected to other nonclassical properties of the field.
Classical states are shown to obey these inequalities and
for the family of centered Gaussian states the direct
connection between violation of Bell-type inequalities and
squeezing is established.
\end{abstract}
\begin{multicols}{2}
The violation of Bell's inequalities is one of the most striking
features of quantum theory~\cite{bell-physics-64}. The testing 
ground for these inequalities has mostly been the states of the
electromagnetic 
field~\cite{clauser-prd-74,aspect-prl-81,ou-prl-88}.
When a state does not obey Bell-type inequalities it definitely has
essential quantum features which cannot be reconciled with the 
classical notions of reality and locality. In most treatments the
Bell-type inequalities are formulated for specific quantum states.
For the electromagnetic field there have been attempts to generalise
the treatment and relate
the violation of Bell-type inequalities with other general nonclassical 
features of the 
states~\cite{reid-pra-86,anu-pra-91,chubarov-pla-85,brif-pra-98}.

We develop the machinery for analysing the violation of
Bell type inequalities for a general state of the 4-mode
radiation field in a setup of the type shown in
Figure~1.  For the direction {\boldmath $k\/$}, the two
orthogonal polarisation modes are described by the
annihilation operators $a_1\/$ and $a_2\/$, with $a_3\/$ and
$a_4\/$ being similarly chosen for the direction {\boldmath
$k^{\prime}$}. Without any loss of generality we choose
{\boldmath $k$} and {\boldmath $k^{\prime}$} to be in the
plane of the paper. This allows a simple choice for the
directions  $x\/$, $x^\prime\/$ to be in the same
plane while $y\/$ and $y^\prime$ point out of this plane. 
The passive, total photon number conserving, canonical
transformations (which will play an important role in our
analysis) amount to replacing the $a_j\/$'s by their
complex linear combinations $a_j^{\prime}= U_{jk} a_k$,
with $U\/$ being a unitary matrix belonging to $U(4)\/$.
$P_1\/$ and $P_2\/$ are polarisers placed at angles
$\theta_1\/$ and $\theta_2\/$ with respect to the $x\/$
and $x^{\prime}\/$ axes while $D_1\/$ and $D_2\/$ are 
photon detectors.
  
Usually states with strictly one photon in each direction
are considered for violation of Bell-type inequalities; a
general state however could have an {\em arbitrary number
of photons}, and could even be a {\em mixed state}. To
handle such states one needs to generalise the concept of
coincidence counts, stipulate the polariser action on
general quantum states and identify precisely the hermitian
operators for which a hidden variable description is being
assumed. As a result of this generalisation we will show
that a classical state in the quantum optical sense always
obeys these inequalities while a nonclassical state may
violate them, possibly after a passive $U(4)\/$ transformation.
Starting with a general nonclassical state, we subject it
to a general unitary evolution corresponding to passive
canonical transformations $U(4)\/$ before we look for the
violation of Bell-type inequalities.  
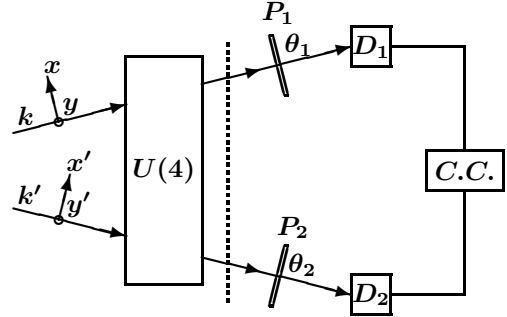
\begin{figure}
\vspace*{1cm}
\hspace*{4.5cm}
\begin{picture}(0,0)(150,100)
\thicklines
\boldmath
\unitlength=0.5truemm
\put(50,10){\framebox(20,60){$U(4)$}}
\multiput(77,5)(0,2){35}{\line(0,1){1}}
\put(20,30){\vector(4,-1){30}}
\put(20,50){\vector(4,1){30}}
\put(70,17){\vector(4,-1){40}}
\put(70,63){\vector(4,1){40}}
\put(86,13){\vector(4,-1){0}}
\put(86,67){\vector(4,1){0}}
\put(90,12){\line(1,4){2}}
\put(90,68){\line(-1,4){2}}
\put(91,12){\line(1,4){2}}
\put(91,68){\line(-1,4){2}}
\put(90,12){\line(-1,-4){2}}
\put(90,68){\line(1,-4){2}}
\put(91,12){\line(-1,-4){2}}
\put(91,68){\line(1,-4){2}}
\multiput(88,4)(0,72){2}{\line(1,0){1}}
\multiput(92,20)(0,40){2}{\line(1,0){1}}
\put(86,80){$P_1$}
\put(90,23){$P_2$}
\put(92,71.5){$\theta_1$}
\put(93,13){$\theta_2$}
\put(110,2){\framebox(10,10){$D_2$}}
\put(110,68){\framebox(10,10){$D_1$}}
\put(120,7){\line(1,0){20}}
\put(120,73){\line(1,0){20}}
\multiput(140,7)(0,38){2}{\line(0,1){28}}
\put(130,35){\framebox(20,10){$C.C.$}}
\put(32,27){\vector(1,4){3}}
\put(32,53){\vector(-1,4){3}}
\put(32,27){\circle{2}}
\put(32,53){\circle{2}}
\put(34,29){$y^{\prime}$}
\put(33,56){$y$}
\put(34,40){$x^{\prime}$}
\put(28,66){$x$}
\put(21,52){\boldmath $k$}
\put(21,31){\boldmath $k^{\prime}$}
\end{picture}
\vspace*{4cm}
\caption{
\narrowtext
Setup to study the violation of Bell
type inequalities for arbitrary states of the four mode
radiation field.}
\end{figure}

A coincidence is defined to occur when both the detectors $D_1\/$
and $D_2\/$ click simultaneously \ie {\em one or more
photons} are detected by each.  The following 
coincidence count rates are considered:

\begin{tabular}{clcl}
(a) &
$P(\theta_1,\theta_2)\/$&~:~~& $P_1\/$ at 
$\theta_1\/$ and $P_2\/$ at $\theta_2$.\\ 
(b)&
$P(\theta_1,\quad\!)\/$&~:~~& $P_2\/$ at
$\theta_1\/$ and $P_2\/$  removed.\\ 
(c)&
$P(\quad\!,\theta_2)\/$&~:~~&$P_1\/$ removed and 
$P_2\/$ at $\theta_2\/$.\\
(d)&
$P(\quad\!,\quad\!)\/$&~:~~& Both 
$P_1\/$ and $P_2\/$ removed.
\end{tabular}

Before further analysis and calculation of these count
rates we need to specify the precise 
of the polarisers on a given quantum state.  Classically,
the action of a polariser is straightforward.  The
component of the electric field along the axis passes
through unaffected while the orthogonal component is
completely absorbed. The quantum action of the polariser is
more complicated: for a given two-mode density matrix
$\rho\/$ (the two polarisation modes for a fixed direction)
incident on a polariser placed at an angle $\theta
\/$, the output(single-mode) state $\rho(\theta)\/$ 
is obtained by taking the trace over the mode orthogonal to
the linear polarisation defined by $\theta\/$. Explicitly
in the number state basis:
\begin{equation}
\rho(\theta) = 
\sum \limits_{n=0}^\infty 
{}_{(\theta+\frac{\pi}{2})}\langle n\vert \rho
\vert n \rangle_{(\theta+\frac{\pi}{2})}
\label{pol-action} 
\end{equation}
The output state $\rho(\theta)\/$ is in general {\em mixed}
even when the input $\rho\/$ is pure.  For the special case
when the input state is a two-mode coherent state the output
single-mode state is once again a coherent state. This is
due to the fact that coherent states are not entangled in
any basis:
\begin{eqnarray}
\vert z_1 \rangle_x \vert z_2 \rangle_y& =&
\vert z_1\cos{\theta}-z_2 \sin{\theta} \rangle_{\theta}
\vert z_1\sin{\theta}+z_2 \cos{\theta}
\rangle_{\theta+\frac{\pi}{2}}\nonumber \\ 
&&\longrightarrow \vert z_1 \cos{\theta}-z_2 \cos{\theta}
\rangle_{\theta}  
\end{eqnarray}
On the other hand, single photon states can in general be
entangled states of the two-mode field and thus would lead
to mixed one-mode states after passing through the
polariser. For example a pure two-mode single photon state
$\frac{1}{\sqrt{2}}(\vert 1
\rangle_x \vert 0\rangle_y+\vert 0
\rangle_x \vert 1 \rangle_y)\/$, after passage through a
polariser placed in the $x\/$ direction reduces to a mixed
state with density matrix $\frac{1}{2}(\vert 0 \rangle_x \,
{}_x \langle 0\vert +
\vert 1 \rangle_x \,{}_x \langle 1\vert)$. 
For a comparable discussion on the action of a beam
splitter see~\cite{campos-pra-89}.

We define the following hermitian operators, with eigen 
values $0\/$ and $1\/$ 
\begin{eqnarray}
\widehat{A}_1 &=& \left( I_{2\times2} - \vert 0 0 \rangle
\langle 0 0 \vert \right)_{\bf k} \nonumber \\
\widehat{A}_2 &=& \left( I_{2\times2} - \vert 0 0 \rangle
\langle 0 0 \vert \right)_{{\bf k}^{\prime}} \nonumber \\
\widehat{A}_1(\theta_1) &=& \left( I_{\theta_1} - 
\vert 0  \rangle_{\!\theta_1}
\,{}_{\theta_1\!}\langle 0 \vert \right)
I_{\theta_1+\frac{\pi}{2}} \nonumber \\
\widehat{A}_2(\theta_2) &=& \left( I_{\theta_2} - 
\vert 0  \rangle_{\!\theta_2}
\,{}_{\theta_2\!}\langle 0 \vert \right)
I_{\theta_2+\frac{\pi}{2}} 
\label{A-operators}
\end{eqnarray}
The subscripts $\theta_1\/$ and $\theta_2\/$ in the last
two equations refer to the settings of $P_1\/$ and $P_2\/$.
$I_{2\times2}\/$ is the two-mode unit operator while
$I_{\theta_{1(2)}}\/$ and $I_{\theta_{1(2)} +
\frac{\pi}{2}}\/$ are one-mode unit operators for the
relevant polarisation modes along the propagation
directions ${\bf k}\/$ or ${\bf k}^{\prime}$.  The
expectation values of these operators are the probabilities of
detecting {\em at least one photon} of the appropriate
kind(For example $\langle \widehat{A}_1 \rangle\/$ is the probability
of detecting at least one photon at $D_1\/$ with $P_1\/$
removed, and $\langle \widehat{A}_1(\theta_1) \rangle$ that
at $D_1\/$ with $P_1\/$ set at $\theta_1\/$).

The quantum mechanical predictions for various coincidence
count rates are the expectation values of the products of pairs
of these operators:
\begin{eqnarray}
P(\theta_1,\theta_2)&=&\langle 
\widehat{A}_1(\theta_1)\;
\widehat{A}_2(\theta_2)
\rangle
\nonumber\\
P(\theta_1,\quad)&=&\langle 
\widehat{A}_1(\theta_1)\;\widehat{A}_2
\rangle
\nonumber \\
P(\quad,\theta_2)&=&\langle 
\widehat{A}_1\;\widehat{A}_2(\theta_2)
\rangle
\nonumber \\
P(\quad,\quad)&=&\langle \widehat{A}_1\;\widehat{A}_2
\rangle
\end{eqnarray}
We note here that due to the definitions~(\ref{A-operators})
the polariser action~(\ref{pol-action}) is automatically
implemented!

In the case when a hidden variable theory is assumed the
value of the hidden variable along with the state vector
will give us the actual outcomes of the individual
measurements for the dynamical variables $A_1, A_2,
A_1(\theta_1)\/$ and $A_2(\theta_2)\/$. The locality
condition of ``no action at a distance'' can then be
readily used to calculate the coincidence count rates.
Further, these rates are constrained by the following
inequality due to Clauser and Horne~\cite{clauser-prd-74}
\begin{eqnarray}
-P(\quad,\quad)\/&\leq\/&P(\theta_1,\theta_2)
-P(\theta_1,\theta_2^{\prime})
+P(\theta_1^{\prime},\theta_2)
\nonumber \\
&&+P(\theta_1^{\prime},\theta_2^{\prime})
-P(\theta^{\prime}_1,\quad)-P(\quad,\theta_2)
 \/\leq\/ 0
\label{chs-ineq}
\end{eqnarray} 
This is the required generalised Bell-type inequality
relevant for arbitrary multi-photon states. We emphasis
here that the coincidence count rates for general states
have a different meaning as opposed to two-photon states.
For two-photon states, extensively studied in the
literature~\cite{ou-prl-88}(for example the state
$\frac{1}{2} (a^{\dagger}_{1}-a^{\dagger}_{3})
(a^{\dagger}_{4}-a^{\dagger}_{2})
\vert 0 \rangle_{1}\vert 0 \rangle_{2}
\vert 0 \rangle_{3}\vert 0 \rangle_{4}$),
our formalism reduces to the usual one. More explicitly,
the single hermitian operator $\widehat{A}=I-\vert 0
\rangle \langle 0 \vert \/$ giving the probability 
for finding one or more photons reduces effectively to
$a^{\dagger}a\/$. The simplifying relations $P(\theta_1,
\quad) = P(\theta_1, \theta_2)+ P(\theta_1,
\theta_2+ \frac{\pi}{2})\/$ etc. are also obtained from
the reduction of $\widehat{A}$'s and are not valid for
general states.

We now turn to the analysis of interesting multiphoton 
states. Consider the 4-mode coherent states:
\begin{equation}
\vert\/\mbox{\boldmath $z$}\/ \rangle =
\exp (-\displayfrac{1}{2}
\displaystyle \mbox{\boldmath $z$}^{\mbox{\tiny T}}\mbox{\boldmath
$z$}^{\star}) \;
\exp\;{\displaystyle  
(\sum_{j=1}^{4}z_{j} 
a^{\dagger}_{j}}) \vert \mbox{\boldmath $0$}\; \rangle
\end{equation}
where 
$\mbox{\boldmath $z$}^{\mbox{\tiny T}}=
\left(z_{1}\; z_{2}\; z_{3}\; z_{4}\right)\/$ 
is a complex row vector.  The quantum mechanical values of
the coincidence count rates for this case can be computed
quite easily:
\begin{eqnarray}
P(\theta_1,\theta_2) &=& (1-e^{\displaystyle 
-\vert z^{\prime}_{1}\vert^2}\;)
                          (1-e^{\displaystyle 
-\vert z^{\prime}_{3}\vert^2}\;)
\nonumber \\
P(\theta_1,\quad) &=& 
(1-e^{\displaystyle -\vert z^{\prime}_{1}\vert^2}\;)
(1-e^{\displaystyle  -\vert z_{3}\vert^2 -
\vert z_{4}\vert^2}\;)
\nonumber \\
P(\quad,\theta_2) &=& 
(1-e^{\displaystyle -\vert z_{1}\vert^2 -
\vert z_{2}\vert^2}\;)
(1-e^{\displaystyle -\vert z^{\prime}_{3}\vert^2})
\nonumber \\
P(\quad,\quad) &=&
(1-e^{\displaystyle 
-\vert z_{1}\vert^2 -\vert z_{2}\vert^2}\;)
(1-e^{\displaystyle 
-\vert z_{3}\vert^2 -\vert z_{4}\vert^2}\;)
\nonumber  \\
\left(\begin{array}{c}
z_1^{\prime}\\
z_2^{\prime}
\end{array}\right)
 &=& \left(\begin{array}{cc}
\cos{\theta_1} & -\sin{\theta_1}\\
\sin{\theta_1} & \cos{\theta_1}
\end{array} \right) 
\left(\begin{array}{c}
z_1\\
z_2
\end{array}\right)\nonumber \\
\left(\begin{array}{c}
z_3^{\prime}\\
z_4^{\prime}
\end{array}\right)
 &=& \left(\begin{array}{cc}
\cos{\theta_2} & -\sin{\theta_2}\\
\sin{\theta_2} & \cos{\theta_2}
\end{array} \right) 
\left(\begin{array}{c}
z_3\\
z_4
\end{array}\right)
\end{eqnarray}
A typical count rate $P(\theta_1, \theta_2)\/$ factorises,
with the first factor depending solely on $\theta_1\/$ and
the second on $\theta_2\/$. This is a consequence of the
unentangled nature of coherent states and is sufficient to
establish their nonviolation of the Bell-type
inequalities~(\ref{chs-ineq}).

An arbitrary state $\rho\/$ of the 4-mode radiation field
can be expressed in terms of projections onto coherent
states~\cite{klauder}:
\begin{eqnarray}
\rho= 
\displayfrac{1}{\pi^4}\int \varphi(\mbox{\boldmath $z$}) 
\vert \mbox{\boldmath $z$}\rangle
\langle \mbox{\boldmath $z$}\vert 
d^8 \mbox{\boldmath $z$},\quad
\displayfrac{1}{\pi^4}
\int \varphi(\mbox{\boldmath $z$}) 
d^8\mbox{\boldmath $z$}=1
\label{def-diag-coh}
\end{eqnarray}
In quantum optics the diagonal coherent state distribution
function $\varphi(\mbox{\boldmath $z$})\/$ describing the
state $\rho\/$ is used to distinguish between classical and
nonclassical states~\cite{walls-nature-79}. The states with
nonnegative nonsingular $\varphi(\mbox{\boldmath $z$})\/$
are classical while the ones with negative or singular(worse
than a delta function) $\varphi(\mbox{\boldmath $z$})\/$
are nonclassical.

The function $\varphi\/$ undergoes a point transformation
when the state undergoes a unitary evolution corresponding
to a passive canonical transformation given by an element of
$U(4)$:
\begin{equation} 
\varphi(\mbox{\boldmath $z$}) \rightarrow
\varphi^{\prime}(\mbox{\boldmath $z$})=
\varphi(\mbox{\boldmath $z$}^{\prime}),\quad 
\mbox{\boldmath $z$}^{\prime}= U\;\mbox{\boldmath $z$},
\;U \in U(4).
\end{equation}
Thus, the classical or nonclassical nature of a state is 
preserved under such transformations.

In principle, $\varphi(\mbox{\boldmath $z$})\/$ can be used
to calculate coincidence count rates for any given state.
In particular for {\em classical states}, they are just
their coherent state values integrated over the positive
distribution function $\varphi(\mbox{\boldmath $z$})\/$.
When such count rates are substituted in the Bell-type
inequality~(\ref{chs-ineq}) it becomes the inequality for
coherent states integrated over a normalized positive
$\varphi(\mbox{\boldmath $z$})\/$. Since coherent states
obey this inequality the integration over such a
distribution obviously preserves this property.  Thus we
conclude that a ``classical state'' will not violate the
Bell type inequalities~(\ref{chs-ineq}). Since the
classical or nonclassical status of a 4-mode state is
invariant under $U(4)\/$, the group of passive canonical
transformations, a classical state after undergoing such
a transformation will still not violate Bell type
inequalities. On the other hand, the nonclassical states
can violate these inequalities; in fact, the violation of
such an inequality implies that the underlying
$\varphi(\mbox{\boldmath $z$})\/$ for the state is negative
or singular and the state is nonclassical in the quantum
optical sense.

The family of squeezed thermal states, which are in general
mixed states and possess a fluctuating number of photons 
vividly illustrate the strength of our formalism.  Consider
a 4-mode state with a centered Gaussian Wigner
distribution~\cite{arvind-pramana-95}
\begin{eqnarray}
W(\xi)&=&\pi^{-4} ({\rm Det} G)^{\frac{1}{2}} 
\exp(-\xi^{T} G \xi), 
\nonumber \\
\xi^{T}&=&\left(
\begin{array}{cccccccc}
q_{1}& q_{2}&q_{3}&
q_{4}&p_{1}&p_{2}&
p_{3}&p_{4}
\end{array}\right)
\nonumber\\ 
G &=& G^*=G^T 
\nonumber \\
G^{-1} + i \beta &\geq& 0, \quad \beta=
\left( \begin{array}{cc}
 0_{4\times4} & 1_{4\times4}\\
-1_{4\times4} & 0_{4\times4}
\end{array} \right)
\label{4-mode-gauss}
\end{eqnarray}
Here, $q_{1}=\frac{1}{\sqrt{2}}(a_{1}^{\dagger}+a_{1}),
p_{1}=\frac{i}{\sqrt{2}}(a_{1}^{\dagger}-a_{1})$ etc. are
the quadrature components. The matrix
$V=\frac{1}{2}G^{-1}\/$ is the variance or the noise
matrix.  For a given state, if the smallest eigenvalue of
this matrix is less than $\frac{1}{2}\/$ then the state is
squeezed and therefore nonclassical~\cite{simon-pra-94}.

We now look at specific examples of such Gaussian 
states in order to illustrate their
violation of Bell-type inequalities. Take
\begin{eqnarray}
G = U^{-1} S^T G_{0} S U,\quad
G_{0}= \kappa \/ I_{8\times 8},\nonumber \\
\quad 0 \leq 
\kappa \leq 1,\quad 
\kappa= 
\tanh \displayfrac{\hbar \omega}{2 k T}.
\end{eqnarray}
Here $\kappa=1\/$ implies zero temperature and $ \kappa <
1\/$ corresponds to some finite temperature, $S\/$ is a
4-mode squeezing symplectic transformation, an
$Sp(8,\Re)\/$ matrix, and $U\/$ is a passive symplectic
$U(4)\/$ transformation whose role is to produce
entanglement. As an example, we start with a state in which
the modes $a_{1}\/$ and $a_{4}\/$ are squeezed by equal and
opposite amounts $u\/$ and the modes $a_{2}\/$ and
$a_{3}\/$ are squeezed by equal and opposite amounts $v\/$,
and the entanglement is ``maximum''. This corresponds to
the choices
\begin{eqnarray}
S={\rm Diag}\left(\begin{array}{cccccccc}
e^{-u}, &e^{v}, &e^{-v}, &e^{u}, &e^{u}, &e^{-v}, &e^{v}, &e^{-u}
\end{array}
   \right), \nonumber \\
U=\displayfrac{1}{2}\left[\begin{array}{cc}
X&0\\
0&X\end{array}\right], \; 
X=\left[\begin{array}{cc} 
  Y&Y\\-Y&Y\end{array}\right],\; 
Y=\left[\begin{array}{cc} 
  1&1\\-1&1\end{array}\right]. 
\end{eqnarray}
For this particular class of centered Gaussian Wigner
states the function
\begin{eqnarray}
f(\theta_1,\theta_2, \theta_1^{\prime}, \theta_2^{\prime})
&&=
P(\theta_1,\theta_2)-
P(\theta_1,\theta^{\prime}_2)+
P(\theta^{\prime}_1,\theta_2)+
\nonumber \\
&&P(\theta^{\prime}_1,\theta^{\prime}_2)
-P(\theta^{\prime}_1,\quad)-
P(\quad,\theta_2)
\end{eqnarray}
can be calculated. 
\begin{figure}
\psfig{figure=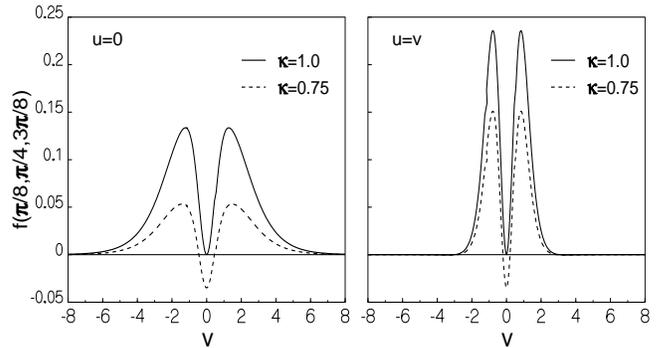,height=5.5cm,width=9cm,angle=0}
\caption{Violation of Bell type inequality for
states with centered Gaussian Wigner distributions
representing a 4-mode centered Gaussian states.} 
\end{figure}
Though one ought to search over all values of the angles
$\theta_1, \theta_2, \theta_1^{\prime}\/$ and
$\theta^{\prime}_2\/$ to locate possible violations of
the inequality~(\ref{chs-ineq}), we choose special values
for the angles and demonstrate the violation.  Plots for
the function $f(\displayfrac{\pi}{8},\displayfrac{\pi}{4}
,3\displayfrac{\pi}{8},0)\/$ are shown in Figure~2 clearly
demonstrating that the Bell-type inequalities are violated
by these states. The solid lines represent the squeezed
vacuum while the dotted lines are for the case of finite 
temperature. We note that even at finite temperatures the 
inequalities are violated and their violation stems from
the nonclassical property of squeezing.

A pure quantum mechanical state of a composite system is
said to be entangled if we are not able to express it as a
product of two factors.  Such states have nontrivial
quantum correlations and can lead to the violation of
suitable Bell-type inequalities.  The passive canonical
transformations $U(4)\/$ have been used to manipulate
entanglement properties of 4-mode states. This capacity
of passive transformations can already be seen at the level
of two-mode fields.  The group of passive canonical
transformations in this case is $U(2)\/$; its elements,
though incapable of producing or destroying nonclassicality
are capable of entangling(disentangling) originally
unentangled(entangled) states.  For example the unentangled
nonclassical state $\vert 1 \rangle \vert 1
\rangle \/$ becomes the  entangled state 
$\frac{1}{\sqrt{2}}(\vert 2
\rangle
\vert 0 \rangle + \vert 0 \rangle \vert 2 \rangle)\/$
by the $U(2)\/$ 
transformation $\exp [(i\pi/4) (a_1^{\dagger}a_2+ a_2^{\dagger}a_1)]\/$. 
However, coherent
states are not entanglable in this way!  Classical states
are statistical mixtures of coherent states and under
$U(2)\/$ remain classical. Such a mixture can definitely
have correlations which are purely classical, but it cannot
have truly quantum mechanical entanglement.
Thus classical states are  to be regarded as nonentangled, 
and they remain
so under passive $U(2)\/$ transformations.  However this is
in general not true for a nonclassical nonentangled state
which may get entangled under a suitable $U(2)\/$
transformation. It is a straightforward matter to
generalise the above statements to $n\/$ mode systems where
the group of passive canonical transformations is $U(n)\/$.

The above conclusions have an interesting bearing on the
work on violation of Bell-type inequalities with beams
originating from independent
sources~\cite{yurke-prl-92,yurke-pra-92,zukowski-prl-92}.
These experiments take two beams from two independent
sources, pass them through some passive optical elements
and show that the Bell-type inequalities are violated. The
first conclusion we can draw from our analysis is that it
must be the quantum optical nonclassicality of one of the
beams in this experiment which has been converted into
entanglement by the $U(4)\/$ transformation and hence led
to the violation. Secondly, if the original beams were
quantum optically classical, no matter what one does, no
violation would be seen. 

In our analysis, we have not distinguished between
strengths of coincidences. The coincidence counter
registers a count when simultaneously each detector detects
one or more photons. This is the reason why we chose the
operators $A$'s to have eigen values $0\/$ and $1\/$.  In
this sense, the measurements involved here are not refined.
It would be interesting to further generalise the analysis
by considering somewhat refined measurements where to some
extent coincidences are
distinguished on the basis of their strengths. However, the
relevant operators in this context may be unbounded; and it
is well known that the formulation of Bell type
inequalities for such operators, though desirable, is
nontrivial.

We have compared quantum optical nonclassicality with
violation of Bell's inequalities. When a state is
nonclassical in the quantum optical sense, it does not
allow a classical description based on an ensemble of
solutions of Maxwell's equations, which is a very specific
classical theory. On the other hand violation by a state of
a Bell type inequality rules out any possibility of
describing it by any general local ``classical'' hidden
variable theory. Therefore, it is understandable that
quantum optical nonclassicality is a necessary but not a
sufficient condition for the violation of Bell's
inequalities. This disparity is partially compensated for
by the freedom to perform passive canonical transformations
on a nonclassical state before looking for violation of
Bell's inequalities though it is not obvious whether this
freedom completely removes this discrepancy. On the other
hand, if a state
obeys Bell's inequalities, it may still not allow a 
``classical'' 
description. Therefore, we need a complete set of Bell's
inequalities capturing the full content of the locality
assumption. These and related aspects will be explored 
elsewhere. 
\leftline{\large \bf Acknowledgement:}
One of the authors (Arvind)
thanks Center for Theoretical Studies, I.I.Sc., Bangalore for
supporting his visit to complete this work.

\end{multicols}

\begin{references}
\bibitem[\dagger]{email}
Email:arvind@physics.iisc.ernet.in
%
\bibitem[\star]{jakkur}
Also at Jawaharlal Nehru Centre for 
Advanced Scientific Research, Jakkur,
Bangalore - 560 064,India.
%
\bibitem{bell-physics-64}
J.~S. Bell,
\newblock {\em Physics}, {\bf 1} 195(1964).
%
%
\bibitem{clauser-prd-74}
J.~F. Clauser and M.~A. Horne.
\newblock {\em Phys. Rev.}, {\bf D 10}, 526(1974).
%
\bibitem{aspect-prl-81}
A.~Aspect, P.~Grangier, and G.~Roger.
\newblock {\em Phys. Rev. Lett.}, {\bf 47}, 460(1981).
%
\bibitem{ou-prl-88}
Z.~Y. Ou and L.~Mandel.
\newblock {\em Phys. Rev. Lett.}, {\bf 61}, 50(1988).
%
\bibitem{reid-pra-86}
M.~D. Reid and D.~F. Walls.
\newblock {\em Phys. Rev.}, {\bf A 34}, 1260(1986).
%
%
%
\bibitem{anu-pra-91}
A.~Venugopalan and R.~Ghosh.
\newblock {\em Phys. Rev.}, {\bf A 44}, 1609(1991).
%
\bibitem{chubarov-pla-85}
M.~S. Chubarov and E.~P. Nikolayev.
\newblock {\em Phys. Lett.}, {\bf A 110}, 199(1985).
%
\bibitem{campos-pra-89}
%
\bibitem{brif-pra-98} C.~Brif, A.~ Mann, M.~Revzen,
\newblock {\em Phys.Rev.} {\bf A 57} 742(1998).
%
R.~A.~Campos, B.~E.~A.~Saleh, and M.~C.~Teich.
\newblock {\em Phys. Rev. Lett.}, {\bf A 40}, 1371(1989).
%
\bibitem{klauder}
J.~R. Klauder and E.~C.~G. Sudarshan.
\newblock Benjamin, New York, 1968.
%
\bibitem{walls-nature-79}
D.~F. Walls.
\newblock {\em Nature}, {\bf 280}, 451(1979).
%
\bibitem{arvind-pramana-95}
Arvind, B.~Dutta, N.~Mukunda, and R.~Simon.
\newblock {\em Pramana Journal of Physics}, {\bf 45}, 471(1995).
%
%
\bibitem{simon-pra-94} R.~Simon, B.~Dutta and N.~Mukunda.
\newblock {\em Phys. Rev.}, {\bf 49}, 1567(1994).
%
\bibitem{yurke-prl-92}
B.~Yurke and D.~Stoler.
\newblock {\em Phys. Rev. Lett.}, {\bf 68}, 1251(1992).
%
\bibitem{yurke-pra-92}
B.~Yurke and D.~Stoler.
\newblock {\em Phys. Rev.}, {\bf A46}, 2229(1992).
%
\bibitem{zukowski-prl-92}
M.~Zukowaski, A. Zeilinger, M.~A.~Horne, and A.~K.~Ekert. 
\newblock {\em Phys. Rev. Lett.}, {\bf 71}, 4287(1992).
\end{references}
\end{document}